\documentclass[floatfix,aps,nofootinbib,onecolumn,12pt]{revtex4}%
\usepackage{amssymb,amsmath}
\usepackage{graphicx}
\usepackage{float}
\usepackage{amsmath}
\usepackage{amsfonts}
\usepackage{amssymb}%
\providecommand{\U}[1]{\protect\rule{.1in}{.1in}}
\newcommand{\ba}{\begin{eqnarray}}
\newenvironment{proof}[1][Proof]{\noindent\textbf{#1.} }{\ \rule{0.5em}{0.5em}}
\newcommand{\bpartial}{\mathop{\partial\kern -4pt\raisebox{.8pt}{$|$}}}
\newcommand{\sbpartial}{\tiny\mathop{\partial\kern -4pt\raisebox{.8pt}{$|$}}}
\newcommand{\bra}{\mathopen{[\kern-1.6pt[}}
\newcommand{\ket}{\mathclose{]\kern-1.5pt]}}
\newcommand{\bbra}{\mathopen{[\kern-2.2pt[\kern-2.3pt[}}
\newcommand{\bket}{\mathclose{]\kern-2.1pt]\kern-2.3pt]}}

\newcommand{\slg}{\mbox{\bfseries\slshape g}}
\newcommand{\sslg}{\mbox{\tiny \bfseries\slshape g}}

\newcommand{\ea}{\end{eqnarray}}
\newcommand{\bege}{\begin{equation}}
\newcommand{\enge}{\end{equation}}
\newcommand{\beq}{\begin{eqnarray}}
\newcommand{\benu}{\begin{enumerate}}
\newcommand{\enu}{\end{enumerate}}
\newcommand{\eeq}{\end{eqnarray}}

\newtheorem{theorem}{Theorem}

\newtheorem{corollary}[theorem]{Corollary}

\newtheorem{lemma}[theorem]{Lemma}

\newtheorem{proposition}[theorem]{Proposition}
\newtheorem{remark}[theorem]{Remark}

\begin{document}
\title{The Maxwell and Navier-Stokes Equations that Follow from Einstein Equation in
a Spacetime Containing a Killing Vector Field}
\author{Fabio Grangeiro Rodrigues}
\email{fabior@ime.unicamp.br}
\affiliation{Institute of Mathematics Statistics and Scientific Computation, IMECC-UNICAMP,
13083-950 Campinas, SP, Brazil}
\author{Waldyr A. Rodrigues Jr.}
\email{walrod@ime.unicamp.br, walrod@mpc.com.br}
\affiliation{Institute of Mathematics Statistics and Scientific Computation, IMECC-UNICAMP,
13083-950 Campinas, SP Brazil}
\author{Rold\~ao da Rocha}
\email{roldao.rocha@ufabc.edu.br}
\affiliation{Centro de Matem\'{a}tica, Computa\c{c}\~{a}o e Cogni\c{c}\~{a}o, Universidade
Federal do ABC, 09210-170, Santo Andr\'{e}, SP, Brazil}
\date{July 01 2012}

\begin{abstract}
In this paper we are concerned to reveal that any spacetime structure $\langle
M,%
\slg
,D,\tau_{%
\sslg
},\nolinebreak\uparrow\rangle$, which is a model of a gravitational field in
General Relativity generated by an energy-momentum tensor $\mathbf{T}$ --- and
which contains at least one nontrivial Killing vector field $\mathbf{A}$ ---
is such that the $2$-form field $F=dA$ (where $A=%
\slg
(\mathbf{A}\boldsymbol{,}$ $)$) satisfies a Maxwell like equation --- with a
well determined current that contains a term of the superconducting type---
which follows directly from Einstein equation. Moreover, we show that the
resulting Maxwell like equations, under an additional condition imposed to the
Killing vector field, may be written as a Navier-Stokes like equation as well.
As a result, we have a set consisting of Einstein, Maxwell and Navier-Stokes
equations, that follows sequentially from the first one under precise
mathematical conditions and once some identifications about field variables
are evinced, as explained in details throughout the text. We compare and
emulate our results with others on the same subject appearing in the
literature. In Appendix A we fix our notation and recall some necessary
material concerning the theory of differential forms, Lie derivatives and the
Clifford bundle formalism used in this paper. Moreover, we comment in Appendix
B on some analogies (and main difference) between our results with ones
obtained long ago by Bergmann and Kommar which are reviewed and briefly criticized.

\end{abstract}
\maketitle

\section{Introduction}

Einstein equations -- relating geometry and matter -- and the Navier-Stokes
equation -- that dictates fluid hydrodynamics -- have been interplaying their
roles recently. Every solution of the incompressible Navier-Stokes equation in
$p + 1$ dimensions was shown to be uniquely associated to a dual solution of
the vacuum Einstein equations in $p + 2$ dimensions.\cite{breke}. Furthermore,
the authors in \cite{min} show that cosmic censorship might be associated to
global existence for Navier- Stokes or the scale separation characterizing
turbulent flows, and in the context of black branes in AdS$_{5}$, Einstein
equations are show to be led to nonlinear equations of boundary fluid dynamics
from \cite{ba2}. In addition, gravity variables can provide a geometrical
framework for investigating fluid dynamics, in a sense of a geometrization of
turbulence \cite{eking}.

Our main aim in this paper is further to provide precisely the formal
relationship between such equations, together with the Maxwell equations. More
specifically, in General Relativity, a Lorentzian spacetime structure (LSTS)
$\langle M,\boldsymbol{g},D,\tau_{\boldsymbol{g}},\uparrow\rangle$ represents
a given gravitational field \cite{sawu}, generated by an energy-momentum
distribution $\mathbf{T}\in\sec T_{0}^{2}M$, which dynamics is determined by
Einstein equation. In this paper we assume \emph{ab initio} that the LSTS
$\langle M,\boldsymbol{g},D,\tau_{\boldsymbol{g}},\uparrow\rangle$ is solely
an effective description of a gravitational field \cite{fr2010,rod2011}, and
that indeed all physical fields are interpreted in the sense of Faraday and
living in a Minkowski spacetime structure $\langle M=\mathbb{R}^{4}%
,\boldsymbol{\mathring{g}},\mathring{D},\tau_{\boldsymbol{\mathring{g}}%
},\uparrow\rangle$. Under this assumption, our main aim in this paper is to
show (Section 2) that when the effective LSTS $\langle M,\boldsymbol{g}%
,D,\tau_{\boldsymbol{g}},\uparrow\rangle$ possesses at least one
\emph{nontrivial} Killing vector field $\mathbf{A\in}\sec TM$, if we denote by
$A=\boldsymbol{g}(\mathbf{A},)\in\sec%
{\textstyle\bigwedge\nolimits^{1}}
T^{\ast}M$ the Killing $1$-form field, then assuming the validity of Einstein
equation, the field $F=dA$ satisfies Maxwell like equations with a well
determined current\footnote{Our proof of this statement in the Clifford bundle
formalism \cite{rodcap2007} is straightforward when compared with standard
tensorial methods.}. Moreover, the Maxwell like equations satisfied by $F$ are
compatible with Einstein equation in a sense explained in the proof of
Proposition 1. We furthermore delve into this approach and prove (Section 3)
that the Maxwell like equations (the one that follows from Einstein equation)
may be written as a Navier-Stokes equation for an inviscid fluid, once we
identify a) a relation between the Killing 1-forms $A$ and\footnote{The
$1$-form fields $A$ and $\mathring{A}$ are related by $A=\mathit{g}%
(\mathring{A})$ \ and \ $F=d\mathit{g}(\mathring{A}$) where $\mathit{g}$ \ is
a well defined extensor field \cite{fr2010,rodcap2007} (see below).}
$\mathring{A}=\boldsymbol{\mathring{g}}(\mathbf{A},)$ and b) the components of
$\mathring{F}=d\mathring{A}$ to some variables which appear in the
Navier-Stokes equation, including as $\boldsymbol{postulate}$ that the
electric like components of $\mathring{F}$ are equal to the components of the
Lamb vector field $\boldsymbol{l}$ of the Navier-Stokes fluid minus a well
determined vector field $\boldsymbol{d}$ that we impose to be a gradient of a
smooth function $\chi$. Obviously, not any Killing vector field $\mathbf{A}$
in the LSTS $\langle M,\boldsymbol{g},D,\tau_{\boldsymbol{g}},\uparrow\rangle$
satisfies that postulate, but as an example in Section III shows, the
postulate has some non trivial realizations. The equation $d\mathring{F}=0$
then becomes the Helmholtz equation for conservation of vorticity. In
addition, since $F=\mathring{F}+G$ (Remark \ref{rdF=0}) where $G$ is a closed
$2$-form field, the homogeneous Maxwell like equation for $F$ is also
equivalent to the Helmholtz equation for conservation of vorticity. In
addition, taking again into account that the Navier-Stokes fluid flows in
Minkowski spacetime structure and that the LSTS structure $\langle
M,\boldsymbol{g},D,\tau_{\boldsymbol{g}},\uparrow\rangle$ is an effective one,
according to ideas developed in \cite{fr2010,rod2011}, the non homogeneous
Maxwell\ like equation for $F$ (and hence the one for $\mathring{F}$ as well)
written in terms of the flat connection $\mathring{D}$ results in a set of
algebraic equations (Eq.(\ref{last})) for the components of $A$ (or for
$\mathring{A}$). Such equations constrain the identification of its components
to the fields appearing in the Navier-Stokes and Einstein equations, since the
energy-momentum tensor of the matter field gets related to the field variables
associated to the Navier-Stokes equation model.

To summarize, we find Maxwell (like) and Navier-Stokes (like) equations that
encodes in precise mathematical sense discussed below the contents of Einstein
equation when some well defined conditions are satisfied for a given arbitrary
nontrivial Killing vector field of the effective LSTS $\langle
M,\boldsymbol{g},D,\tau_{\boldsymbol{g}},\uparrow\rangle$. In Section 4 we
present our conclusions, where our achievements are briefly compared with
other proposals to identify a correspondence between solutions of Einstein and
Navier-Stokes equations in spacetimes of different dimensions.

Also, in Appendix A we fix our notation and recall some necessary material
concerning the theory of differential forms, Lie, derivatives and the Clifford
bundle used in this paper. Moreover, we comment in Appendix B on some
analogies (and main difference) between our results with ones obtained long
ago by Bergmann and Kommar which are reviewed and briefly criticized.

\section{The Maxwell like Equation Following from Einstein Equation}

In this Section we prove two lemmata, a proposition and a corollary whose
objective is to obtain (given the conditions satisfied by $A$ mentioned in the
Introduction) Maxwell like equations for the electromagnetic \emph{like} field
$F=dA$ (where $A=\boldsymbol{g}(\mathbf{A},$ $)$) with a well determined
current like term. More precisely we have

\begin{proposition}
Let $A=\boldsymbol{g}(\mathbf{A},$ $)\in\sec%
{\textstyle\bigwedge\nolimits^{1}}
T^{\ast}M$, where $\mathbf{A}$ is a nontrivial Killing vector field in a LSTS
$\langle M,\boldsymbol{g},D,\tau_{\boldsymbol{g}},\uparrow\rangle$ which
represents a gravitational field generated by\ a given energy-momentum
distribution $\mathbf{T}\in\sec T_{0}^{2}M$ according to Einstein equation.
Then $A$ satisfies the wave equation%
\begin{equation}
\square A-\frac{R}{2}A=\boldsymbol{T}\mathcal{(}A\mathcal{)},\label{1}%
\end{equation}
where $\square$ is the covariant D'Alembertian\emph{\footnote{In this paper we
use the nomenclature and (whenever possible) the notations in
\cite{rodcap2007}. The covariant D'Alembertian is $\square
=\boldsymbol{\partial\cdot\partial}$ where $\boldsymbol{\partial=\vartheta
}^{\mu}D_{\boldsymbol{e}_{\mu}}$ is the Dirac operator acting on sections of
the Clifford bundle $\mathcal{C}\ell(M,g).$}}, $R$ is the scalar curvature,
$\boldsymbol{T}\mathcal{(}A\mathcal{)}:=\mathcal{T}^{\mu}A_{\mu}\in\sec%
{\textstyle\bigwedge\nolimits^{1}}
T^{\ast}M$, where the $\mathcal{T}^{\mu}=T_{\nu}^{\mu}\boldsymbol{\vartheta
}^{\nu}\in\sec%
{\textstyle\bigwedge\nolimits^{1}}
T^{\ast}M$ are the energy-momentum $1$-form fields, with $\mathbf{T}=T_{\mu
\nu}\boldsymbol{\vartheta}^{\mu}\otimes\boldsymbol{\vartheta}^{\nu}$. Moreover
Eq.\emph{(\ref{1}) }is always compatible with Einstein
equation\emph{\footnote{Keep in mind that the validity of Einstein equation is
an hypothesis in the proposition.}}.
\end{proposition}

Moreover, denoting $F=dA\nonumber$ and by $\underset{\boldsymbol{g}}{\delta}$
the Hodge coderivative operator, we have the

\begin{corollary}%
\begin{equation}
dF=0,\qquad\text{ }\underset{\boldsymbol{g}}{\delta}F=-RA-2\boldsymbol{T}%
\mathcal{(}A\mathcal{)}. \label{3}%
\end{equation}

\end{corollary}

Before proceeding, note that the field $F\in\sec%
{\textstyle\bigwedge\nolimits^{2}}
T^{\ast}M$ satisfies Maxwell \emph{like} equations with a current
$J=RA-2\boldsymbol{T}\mathcal{(}A\mathcal{)}$ that splits in a part $J_{s}%
=RA$, of the \textquotedblleft\textit{superconducting\textquotedblright\ }type.

In order to present proofs for the propositions above, the bundle of
differential forms is embedded in the Clifford bundle\footnote{$\mathcal{A}$
$\hookrightarrow\mathcal{B}$ means that $\mathcal{A}$ is embedded in
$\mathcal{B}$ and $\mathcal{A\subseteq B}$.} --- $%
{\textstyle\bigwedge}
T^{\ast}M=%
{\textstyle\bigoplus\nolimits_{r=0}^{4}}
{\textstyle\bigwedge}
T^{r}M\hookrightarrow\mathcal{C}\ell(M,\mathtt{g})$, where $\mathcal{C}%
\ell(M,\mathtt{g})$ is the Clifford bundle of differential forms
\cite{rodcap2007} where $\mathtt{g}$ is the metric of the cotangent bundle.
\cite{mosna}.

To start the proof of the propositions we need two lemmata whose detailed
proofs are given in \cite{rodaaca2010}.

\begin{lemma}
\label{lemma1}If a vector field $\mathbf{A\in}\sec TM$, with $M$ part of the
structure $\langle M,\boldsymbol{g},D,\tau_{\boldsymbol{g}},\uparrow\rangle$
is a Killing vector field, then $\underset{\boldsymbol{g}}{\delta}A=0$, where
$A=\boldsymbol{g}(\mathbf{A},)=A_{\mu}\boldsymbol{\vartheta}^{\mu}=A^{\mu
}\boldsymbol{\vartheta}_{\mu}$.
\end{lemma}

\begin{lemma}
\label{lemma2}If $\mathbf{A\in}\sec TM$ \emph{(}where $M$ is part of the
structure $\langle M,\boldsymbol{g},D,\tau_{\boldsymbol{g}},\uparrow\rangle
$\emph{)} is a Killing vector field, then%
\begin{equation}
\boldsymbol{\partial\wedge\partial}A=\square A=\mathcal{R}^{\mu}A_{\mu},
\label{4}%
\end{equation}
where $\boldsymbol{\partial=\vartheta}^{\mu}D_{e_{\mu}}$ is the Dirac operator
acting on the sections of the Clifford bundle $\mathcal{C}\ell(M,\mathtt{g})$
and $\boldsymbol{\partial\wedge\partial}$ is the Ricci operator acting on
$\sec%
{\textstyle\bigwedge\nolimits^{1}}
T^{\ast}M\hookrightarrow\sec\mathcal{C}\ell(M,\mathtt{g})$. Finally
$\mathcal{R}^{\mu}\in\sec%
{\textstyle\bigwedge\nolimits^{1}}
T^{\ast}M\hookrightarrow\sec\mathcal{C}\ell(M,\mathtt{g})$ are the Ricci
$1$-form fields, with $\mathcal{R}^{\mu}=R_{\nu}^{\mu}\boldsymbol{\vartheta
}^{\nu}$, where $R_{\nu}^{\mu}$ are the components of the Ricci tensor.
\end{lemma}

Now all pre-requisites necessary to prove Proposition 1 are presented, but
before its proof is evinced an observation is necessary.

\begin{remark}
In the theory presented in \emph{\cite{rodaaca2010}} the field $A$ is supposed
to be \emph{(}up to a dimensional constant\emph{)} the electromagnetic
potential of a genuine electromagnetic field created by a given
superconducting current and interacting with the gravitational field. Then,
clearly, the field $F=dA$ in \emph{\cite{rodaaca2010}} is supposed to
automatically satisfy Maxwell equations. In what follows we only suppose that
$A=\boldsymbol{g}(\mathbf{A},)$, where $\mathbf{A}$ is a Killing vector field
in the structure $\langle M,\boldsymbol{g},D,\tau_{\boldsymbol{g}}%
,\uparrow\rangle$.\linebreak\ It is not supposed that $A$ is the
electromagnetic potential of a genuine electromagnetic field.
\end{remark}

\begin{proof}
(of Proposition1) Under the hypothesis of\ Proposition 1, Einstein equation
(in geometrical units) is written as $Ricci-\frac{1}{2}R\boldsymbol{g}%
=\mathbf{T}$ ($Ricci=R_{\mu\nu}\boldsymbol{\vartheta}^{\mu}\otimes
\boldsymbol{\vartheta}^{\nu}$) and can be rewritten in the equivalent form
\cite{landau}%
\begin{equation}
\mathcal{R}^{\mu}-\frac{1}{2}R\boldsymbol{\vartheta}^{\mu}=\mathcal{T}^{\mu}.
\label{5}%
\end{equation}
Now, we use the fact that the Ricci operator satisfies $\boldsymbol{\partial
\wedge}\boldsymbol{\partial\vartheta}^{\mu}=\mathcal{R}^{\mu}$ and moreover
that it is an extensorial operator\footnote{Note that the covariant
D'Alembertian $\ \boldsymbol{\partial\cdot\partial}$ is not an extensorial
operator, i.e., in general $\boldsymbol{\partial\cdot\partial A\neq}A_{\mu
}\boldsymbol{\partial\cdot\partial\vartheta}^{\mu}$.}, i. e., $A_{\mu
}(\boldsymbol{\partial\wedge\partial}\boldsymbol{\vartheta}^{\mu
})=\boldsymbol{\partial\wedge\partial}A$. After multiplying Eq.(\ref{5}) by
$A_{\mu}$ it follows that
\begin{equation}
\boldsymbol{\partial\wedge\partial}A-\frac{1}{2}RA=\boldsymbol{T}%
\mathcal{(}A\mathcal{)}, \label{6}%
\end{equation}
where $\boldsymbol{T}\mathcal{(}A\mathcal{)}=\mathcal{T}^{\mu}A_{\mu}\in\sec%
{\textstyle\bigwedge\nolimits^{1}}
T^{\ast}M\hookrightarrow\sec\mathcal{C}\ell(M,\mathtt{g})$. Now using
Eq.(\ref{4}) which states that $\boldsymbol{\partial\wedge\partial}A=\square
A$, it reads
\begin{equation}
\square A-\frac{1}{2}RA=\boldsymbol{T}\mathcal{(}A\mathcal{)}, \label{6a}%
\end{equation}
which proves the first part of the proposition. To prove that Eq.(1) is
compatible with Einstein equation we start using Eq.(\ref{4}), i.e.,
$\boldsymbol{\partial\wedge\partial}A=\square A=\mathcal{R}^{\mu}A_{\mu}$ .
Eq.(\ref{1}) can be written after some trivial algebra as $(R_{\nu}^{\mu
}-\frac{1}{2}R\delta_{\nu}^{\mu}+T_{\nu}^{\mu})A_{\mu}=0$. Next we observe
that even if some of the $A_{\mu}$ are zero in a given coordinate basis it is
always possible to find a new coordinate basis where all the $A_{\mu}^{\prime
}=\frac{\partial x\prime^{\nu}}{\partial x^{\mu}}A_{\mu}\neq0$. In this new
basis we have%
\begin{equation}
(R_{\nu}^{\prime\mu}-\frac{1}{2}R\delta_{\nu}^{\mu}-T_{\nu}^{\prime\mu}%
)A_{\mu}^{\prime}=0. \label{LHE}%
\end{equation}
Now, Eq. (\ref{LHE}) is a homogeneous system of linear equations for the
variables $A_{\mu}^{\prime}$ and since all the $A_{\mu}^{\prime}\neq0$,\ it is
necessary that $\det(R_{\nu}^{\prime\mu}-\frac{1}{2}R\delta_{\nu}^{\mu}%
-T_{\nu}^{\prime\mu})=0$. It cannot be the case that $R_{\nu}^{\prime\mu
}-\frac{1}{2}R\delta_{\nu}^{\mu}+T_{\nu}^{\prime\mu}\neq0$, for the validity
of Einstein equation is assumed as hypothesis. So, it follows that Eq.(1) is
compatible with the validity of Einstein equation.\smallskip
\end{proof}

\begin{proof}
(of Corollary 2) To prove the corollary we sum $\square A=\boldsymbol{\partial
\cdot\partial}A$ to both members of Eq.(\ref{6}) and take into account that
for any $\mathcal{C\in}\sec\mathcal{C}\ell(M,\mathtt{g})$ the following
expression \cite{rodcap2007} $\boldsymbol{\partial}^{2}\mathcal{C}%
=\boldsymbol{\partial\wedge\partial}\mathcal{C+}\boldsymbol{\partial
\cdot\partial}\mathcal{C}$ holds. Then
\begin{equation}
\boldsymbol{\partial}^{2}A=\frac{1}{2}RA+\boldsymbol{T}\mathcal{(}%
A\mathcal{)}+\square A. \nonumber\label{7}%
\end{equation}
Now, since $\boldsymbol{\partial}^{2}A=-\underset{\boldsymbol{g}}{\delta
}dA-d\underset{\boldsymbol{g}}{\delta}A$, and Lemma \ref{lemma1} implies that
$\underset{\boldsymbol{g}}{\delta}A=0$, it follows that $\boldsymbol{\partial
}^{2}A=-\underset{\boldsymbol{g}}{\delta}F$. Finally, taking into account
Eq.(\ref{4}) it follows that
\begin{equation}
\underset{\boldsymbol{g}}{\delta}F=-\boldsymbol{J}. \label{M1}%
\end{equation}
Of course, $\boldsymbol{J}=-\underset{\boldsymbol{g}}{\delta}%
F=\boldsymbol{\partial}^{2}A=\boldsymbol{\partial\cdot\partial}%
A+\boldsymbol{\partial\wedge\partial}A=2\mathcal{R}^{\mu}A_{\mu}$ which is
equivalent to Eq.(1.7) in Papapetrou paper \cite{papa} obtained in component
form. See also Eq.(5) of \cite{faso}. When Einstein equation is valid we can
immediately write taking into account Eq.(\ref{5})%
\begin{equation}
\boldsymbol{J}=RA+2\boldsymbol{T}\mathcal{(}A\mathcal{)} \label{curr}%
\end{equation}
and the corollary is proved.
\end{proof}

\begin{remark}
We remark that since $\boldsymbol{\partial}=d-\underset{\boldsymbol{g}%
}{\delta}$ we can write a single Maxwell like equation\emph{\footnote{No
misprint here!}} for the field $F$ associated to the Killing form $A$, i.e.,%
\begin{equation}
\boldsymbol{\partial}F=RA+2\boldsymbol{T}\mathcal{(}A\mathcal{)}\text{.}
\label{8}%
\end{equation}

\end{remark}

In \cite{fr2010} it was shown that if the manifold $M$ is
\textit{parallelizable}\footnote{The motivation being Geroch theorem
\cite{geroch} which says that a necessary and sufficient condition for a
4-dimensional Lorentzian manifold $\langle M,%
\slg
\rangle$ to admit spinor fields is that the orthonormal frame bundle be
trivial, which implies that the manifold is parallelizable.}, i. e., there
exists four global vector fields $\boldsymbol{e}_{\mathbf{a}}\in\sec TM$,
$\mathbf{a}=0,1,2,3$ with $\{\boldsymbol{e}_{\mathbf{a}}\}$ a basis for $TM$.
Take $\{\mathfrak{g}^{\mathbf{a}}\}$ as the dual basis of the
$\{\boldsymbol{e}_{\mathbf{a}}\}$. If a LSTS $\langle M,\boldsymbol{g}%
,D,\tau_{\boldsymbol{g}},\uparrow\rangle$ is introduced by postulating that
$\boldsymbol{g}=\eta_{\mathbf{ab}}\mathfrak{g}^{\mathbf{a}}\otimes
\mathfrak{g}^{\mathbf{b}}$, then the gravitational field is described by field
equations --- equivalent in a precise mathematical sense to Einstein equation
--- satisfied by the \textit{potentials} $\mathfrak{g}^{\mathbf{a}}$. In
addition, they are derived through a variational principle from a Lagrangian
density%
\begin{equation}
\mathcal{L}_{g}=\frac{1}{2}d\mathfrak{g}^{\mathbf{a}}\wedge
\underset{\boldsymbol{g}}{\star}d\mathfrak{g}_{\mathbf{a}}-\frac{1}%
{2}\underset{\boldsymbol{g}}{\delta}\mathfrak{g}^{\mathbf{a}}\wedge
\underset{\boldsymbol{g}}{\star}\underset{\boldsymbol{g}}{\delta}%
\mathfrak{g}_{\mathbf{a}}-\frac{1}{4}d\mathfrak{g}^{\mathbf{a}}\wedge
\mathfrak{g}_{\mathbf{a}}\wedge\underset{\boldsymbol{g}}{\star}(d\mathfrak{g}%
^{\mathbf{a}}\wedge\mathfrak{g}_{\mathbf{a}}). \label{g10}%
\end{equation}
which differs from the Einstein Hilbert Lagrangian density ( see
Eq.(\ref{n022})) by an exact differential.

The field equations for the fields $\mathcal{F}^{\mathbf{a}}=d\mathfrak{g}%
^{\mathbf{a}}$ $\in\sec%
{\textstyle\bigwedge\nolimits^{2}}
T^{\ast}M$ are:%
\begin{equation}
d\mathcal{F}^{\mathbf{a}}=0,\qquad\text{\ \ \ }\underset{\boldsymbol{g}%
}{\delta}\mathcal{F}^{\mathbf{a}}=-(\mathfrak{t}^{\mathbf{a}}+\mathbf{T}%
^{\mathbf{a}}), \label{11}%
\end{equation}
where the $\mathbf{T}^{\mathbf{a}}$, as above, are the energy-momentum
$1$-form fields of the matter plus non gravitational fields and $\mathfrak{t}%
^{\mathbf{a}}$ are the energy-momentum the gravitational field and which are
indeed \textit{legitimate tensor objects} since in \cite{rod2011} it has been
proved that they have the very nice and straightforward expression when the
$\underset{\boldsymbol{g}}{\delta}\mathfrak{g}^{\mathbf{a}}=0$, i.e., the
potentials are in the Lorentz gauge\footnote{In \cite{rod2011} it is not
explictly mentioned that Eq.(\ref{12}) is true only in the Lorenz gauge.}
\begin{equation}
\mathfrak{t}^{\mathbf{a}}=(\boldsymbol{\partial\cdot\partial)}\mathfrak{g}%
^{\mathbf{a}}+\frac{1}{2}R\mathfrak{g}^{\mathbf{a}}. \label{12}%
\end{equation}
Under the conditions above, if Eq.(\ref{5}) is rewritten in the orthonormal
cobasis $\{\mathfrak{g}^{\mathbf{a}}\}$ it reads
\[
\mathcal{R}^{\mathbf{a}}-\frac{1}{2}R\mathfrak{g}^{\mathbf{a}}=\mathcal{T}%
^{\mathbf{a}}.
\]
Then, as above, taking into account the definitions of the Ricci, the
covariant D'Alembertian, and the Hodge Laplacian operators, together with
Eq.(\ref{12}) and denoting $\mathfrak{t}(A):=\mathfrak{t}^{\mathbf{a}%
}A_{\mathbf{a}}-A_{\mathbf{a}}(\boldsymbol{\partial\cdot\partial)}%
\mathfrak{g}^{\mathbf{a}}$, the equations of motion of our theory under those
conditions are expressed:%
\begin{equation}
dF=0,\qquad\qquad\underset{\boldsymbol{g}}{\delta}F=-2(\mathfrak{t(}%
A)+\boldsymbol{T}\mathcal{(}A\mathcal{))} \label{13}%
\end{equation}
that can be summarized in a single equation with the use of the Dirac operator
$\boldsymbol{\partial}$ acting on sections of the Clifford bundle:
\begin{equation}
\boldsymbol{\partial}F=2(\mathfrak{t(}A)+\boldsymbol{T}\mathcal{(}%
A\mathcal{))}. \label{MEeq.}%
\end{equation}

\section{From Maxwell Equation to a Navier-Stokes Equation}

In this Section we obtain a Navier-Stokes equation that follows from the
Maxwell\ like equation obtained above (that as just showed above encodes
Einstein equation) once we impose that the electric like components of $F=dA$
satisfy\ Eq.(\ref{lamb}). In order to do that we start with the observation
that the original Navier-Stokes equation describes the non relativistic motion
of a general fluid in Newtonian spacetime. It is not thus adequate to use ---
at least in principle --- a general Lorentzian spacetime structure $\langle
M,\boldsymbol{g},D,\tau_{\boldsymbol{g}},\uparrow\rangle$ to describe a fluid
motion. In fact we want to describe a fluid motion in a background spacetime
such that the fluid medium, together with its dynamics, is equivalent to a
Lorentzian spacetime governed by Einstein equation in the sense described below.

In order to proceed, it was proposed in \cite{fr2010} a theory of the
gravitational field, where gravitation is interpreted as a plastic distortion
of the Lorentz vacuum. In that theory the gravitational field is represented
by a $(1,1)$-extensor field $\boldsymbol{h:}\sec%
{\textstyle\bigwedge\nolimits^{1}}
T^{\ast}M\rightarrow\sec%
{\textstyle\bigwedge\nolimits^{1}}
T^{\ast}M$ living in Minkowski spacetime\footnote{Minkowski spacetime is the
structure $\langle M=\mathbb{R}^{4},\overset{\circ}{%
\slg
},\mathring{D},\tau_{\overset{\circ}{%
\sslg
}},\uparrow\rangle)$, where $\overset{\circ}{%
\slg
}$ is Minkowski metric, $\mathring{D}$ is its Levi-Civita connection, and the
remaining symbols define the spacetime orientation and the time orientation.}.
The field $\boldsymbol{h}$ --- generated by a given energy-momentum
distribution in some region $U$ of Minkowski spacetime --- distorts the
Lorentz vacuum described by the global cobasis\footnote{The $\langle
\mathtt{x}^{\mu}\rangle$ are global coordinate functions in Einstein-Lorentz
Poincar\'{e} gauge for the Minkowskispacetime structure that are naturally
adapted to an inertial reference frame $\mathbf{e}_{0}=\partial/\partial
\mathtt{x}^{0},\mathring{D}\mathbf{e}_{0}=0$ . More details on the concept of
reference frames can be found, e.g., in \cite{rodcap2007,giwal}}
$\langle\boldsymbol{\gamma}^{\mu}=d\mathtt{x}^{\mu}\rangle$, dual with respect
to the basis $\langle\mathbf{e}_{\mu}=\partial/\partial\mathtt{x}^{\mu}%
\rangle$ of $TM$, thus generating the gravitational potentials $\mathfrak{g}%
^{\mathbf{a}}=\boldsymbol{h}(\delta_{\mu}^{\mathbf{a}}\boldsymbol{\gamma}%
^{\mu})$.

Now, in the inertial reference frame $\mathbf{e}_{0}=\partial/\partial
\mathtt{x}^{0}$ (according to the Minkowski spacetime structure $\langle
M=\mathbb{R}^{4},\boldsymbol{\mathring{g}},\mathring{D},\tau
_{\boldsymbol{\mathring{g}}},\uparrow\rangle$), we write using the global
coordinate functions $\langle\mathtt{x}^{\mu}\rangle$ for $M\simeq
\mathbb{R}^{4}$,\footnote{The basis $\{\mathbf{e}^{\mu}\}$ is the
\emph{reciprocal }basis of the basis $\{\mathbf{e}_{\mu}\}$, i.e.,
$\boldsymbol{\mathring{g}}(\mathbf{e}^{\mu},\mathbf{e}_{\nu})=\delta_{\nu
}^{\mu}.$}%
\begin{equation}
\mathbf{A=}\mathring{A}^{\mu}\mathbf{e}_{\mu}:=\left(  \frac{1}{\sqrt
{1-\boldsymbol{v}^{2}}}+V_{0}+q\right)  \mathbf{e}_{0}-v^{i}\mathbf{e}_{_{i}%
}=\mathring{A}_{\mu}\mathbf{e}^{\mu}=\phi\mathbf{e}^{0}-v_{i}\mathbf{e}^{i},
\label{A}%
\end{equation}
where the vector function $\boldsymbol{v}=(v_{1,}v_{2},v_{3})$ is to be
identified with the $3$-velocity of a Navier-Stokes fluid --- in the inertial
frame $\mathbf{e}_{0}$ according to the conditions disclosed below. Also,
$V_{0}$ denotes a scalar function representing an external potential acting on
the fluid, and
\begin{equation}
q=\int_{0}^{(t,\mathbf{x)}}\frac{dp}{\rho}, \label{p}%
\end{equation}
where the functions $p$ and $\rho$ are identified respectively with the
pressure and density of the fluid and supposed functionally related, i.e.,
$dp\wedge dq=0$. Furthermore $\boldsymbol{v}^{2}:=%
{\textstyle\sum\nolimits_{i=1}^{3}}
(v_{i})^{2}$.

Before proceeding note that $\phi$ looks like the relativistic energy per unit
mass of the fluid. Then we will write $\mathbf{A}$ as%
\begin{equation}
\mathbf{A=}\left(  \frac{1}{2}\boldsymbol{v}^{2}+V+q\right)  \mathbf{e}%
_{0}-v^{i}\mathbf{e}_{_{i}}=\phi\mathbf{e}^{0}-v_{i}\mathbf{e}^{i}, \label{FI}%
\end{equation}
where the new potential function $V$ is the sum of $V_{0}$ with the sum of the
Taylor expansion terms of $[(1-\boldsymbol{v}^{2})^{-1/2}-\frac{1}%
{2}\boldsymbol{v}^{2}].$

Then we have%
\begin{equation}
\mathring{A}=\boldsymbol{\mathring{g}}(\mathbf{A,})=\mathring{A}_{\mu
}\boldsymbol{\gamma}^{\mu}=\eta_{\mu\nu}\mathring{A}^{\nu}\boldsymbol{\gamma
}^{\mu}=\mathring{A}^{\mu}\boldsymbol{\gamma}_{\mu},\text{ \ \ }\mathring
{F}=d\mathring{A}=\frac{1}{2}\mathring{F}_{\mu\nu}\boldsymbol{\gamma}^{\mu
}\wedge\boldsymbol{\gamma}^{\nu}, \label{aponto}%
\end{equation}

and%
\begin{equation}
A=\boldsymbol{g}(\mathbf{A,})=A_{\mu}\boldsymbol{\gamma}^{\mu}=g_{\mu\nu
}\mathring{A}^{\nu}\boldsymbol{\gamma}^{\mu}=A^{\mu}\boldsymbol{\gamma}_{\mu
},\text{ \ \ }F=dA=\frac{1}{2}F_{\mu\nu}\boldsymbol{\gamma}^{\mu}%
\wedge\boldsymbol{\gamma}^{\nu}, \label{a}%
\end{equation}

We. proceed by identifying the magnetic like and the electric like components
of the field $\mathring{F}$ with components of the vorticity field and the
Lamb vector field of the fluid. We remark that whereas the identification of
the magnetic like components of $\mathring{F}$ with the vorticity field is
natural, the identification of the electric like components of $F$ is here
$\boldsymbol{postulated}$, i.e. we suppose that $\mathring{F}$ besides being
derived from a Killing vector field associated to the structure also satisfies
the constraint given by Eq.(\ref{lamb}) below. We thus write $\mathring
{F}_{\mu\nu}=(d\mathring{A})_{\mu\nu}$ as
\begin{equation}
\mathring{F}_{\mu\nu}=\left(
\begin{array}
[c]{cccc}%
0 & l_{1}-d_{1} & l_{2}-d_{2} & l_{3}-d_{3}\\
-l_{1}+d_{1} & 0 & -w_{3} & w_{2}\\
-l_{2}+d_{2} & w_{3} & 0 & -w_{1}\\
-l_{3}+d_{3} & -w_{2} & w_{1} & 0
\end{array}
\right)  \label{f}%
\end{equation}
were $\boldsymbol{w}$ is vorticity of the velocity field
\begin{equation}
\boldsymbol{w}:=\nabla\times\boldsymbol{v,} \label{vort}%
\end{equation}
and
\begin{equation}
\boldsymbol{l}:=\boldsymbol{w}\times\boldsymbol{v}, \label{lamb}%
\end{equation}
is the so called \textit{Lamb} vector and moreover%
\begin{equation}
\boldsymbol{d}=-\nabla\chi, \label{d}%
\end{equation}
where $\chi$ is a smooth function.

\begin{remark}
Before proceeding it is important to emphasize that the identification of the
components of $\mathring{F}$ with the components of the Lamb and vorticity
fields has been done in an arbitrary but fixed inertial frame $\mathbf{e}%
_{0}=\partial/\partial\mathtt{x}^{0}$ as introduced above.
\end{remark}

At this point we recall that the non relativistic Navier-Stokes equation for
an \emph{inviscid fluid} is given by \cite{cm,flanders}
\begin{equation}
\frac{\partial\boldsymbol{v}}{\partial t}+(\boldsymbol{v}\cdot\nabla
)\boldsymbol{v}=-\nabla(V+q), \label{NS1}%
\end{equation}
or using a well known vector identity,
\begin{equation}
\frac{\partial\boldsymbol{v}}{\partial t}=-\boldsymbol{w}\times\boldsymbol{v}%
-\nabla\left(  V+\frac{p}{\rho}+\frac{1}{2}\boldsymbol{v}^{2}\right)  .
\label{NS2}%
\end{equation}
By these identifications\footnote{Other identifications of Navier-Stokes
equation with Maxwell equations may be found in \cite{suhan,sudjhan}.}, we get
a Navier-Stokes like equation from the straightforward identification of
$\boldsymbol{l}-\boldsymbol{d}=(\mathring{F}_{01},\mathring{F}_{02}%
,\mathring{F}_{03})$ and $\boldsymbol{w}=(\mathring{F}_{32},\mathring{F}%
_{13},\mathring{F}_{21})$. Indeed, we have
\begin{align}
\mathring{F}_{0i}  &  =(\boldsymbol{w}\times\boldsymbol{v})_{i}-d_{i}%
=-\frac{\partial v_{i}}{\partial t}-\frac{\partial\phi}{\partial\mathtt{x}%
^{i}},\label{NS3a}\\
\mathring{F}_{jk}  &  =-%
{\textstyle\sum\nolimits_{i=1}^{3}}
\epsilon_{ijk}w_{i}, \label{NS3}%
\end{align}
where $\epsilon_{ijk}$ is the $3$-dimensional Kronecker symbol. Eq.(\ref{NS3a}%
) becomes%

\begin{equation}
\frac{\partial\boldsymbol{v}}{\partial t}+\boldsymbol{w}\times\boldsymbol{v}%
+\nabla\left(  \frac{1}{2}\boldsymbol{v}^{2}\right)  =-\nabla\left(
V+\frac{p}{\rho}\right)  +d_{i}, \label{NSE4}%
\end{equation}
and since $\boldsymbol{d}=-\nabla\chi$ for some smooth function $\chi$ then
Eq.(\ref{NSE4}) can be written as%

\begin{equation}
\frac{\partial\boldsymbol{v}}{\partial t}+\boldsymbol{w}\times\boldsymbol{v}%
+\nabla\left(  \frac{1}{2}\boldsymbol{v}^{2}\right)  =-\nabla\left(
V+\chi+\frac{p}{\rho}\right)  \label{nse5}%
\end{equation}
which is now a Navier-Stokes like equation for a fluid moving in an external
potential $V^{\prime}=V+\chi.$

Moreover, the homogeneous Maxwell equation $d\mathring{F}=0$ is equivalent to
\begin{align}
\nabla\times\boldsymbol{l}+\frac{\partial\boldsymbol{w}}{\partial t}  &
=0,\nonumber\\
\nabla\cdot\boldsymbol{w}  &  =0, \label{helm}%
\end{align}
which express Helmholtz equation for conservation of vorticity.

\begin{remark}
\label{rdF=0}Note that since $A=\boldsymbol{g}(\mathbf{A,)=}A_{\mu
}\boldsymbol{\gamma}^{\mu}=\mathring{A}^{\mu}g_{\mu\nu}\boldsymbol{\gamma
}^{\nu}$, we can write\footnote{We have (details in \cite{fr2010})
$\mathit{g}=\boldsymbol{h}^{\dagger}\boldsymbol{h}$ and $\boldsymbol{\mathring
{g}}(\mathit{g}(\boldsymbol{\gamma}_{\mu}),\boldsymbol{\gamma}_{\nu}%
)=g_{\mu\nu}=\boldsymbol{\mathring{g}}(\boldsymbol{h}(\boldsymbol{\gamma}%
_{\mu}),\boldsymbol{h}(\boldsymbol{\gamma}_{\nu}))=\boldsymbol{\mathring{g}%
(}\mathfrak{g}_{\mu},\mathfrak{g}_{\nu})$.}
\[
A=\mathit{g}(\mathring{A}),
\]
where the extensor \ field $\mathit{g}$ is defined by $\mathit{g}%
(\boldsymbol{\gamma}_{\mu})=g_{\mu\nu}\boldsymbol{\gamma}^{\nu}$. Thus, in
general we can write since $d(F-\mathring{F})=0$ , $F=\mathring{F}+G$ where
$G$ is a closed $2$-form field\footnote{Even more, taking into account that
the Minkowski manifold is star shape we have that $G$ is exact. Thus
$F=\mathring{F}+dK$, for some smooth $1$-form field $K$.}. So, $dF=d\mathring
{F}=0$ express the same content, namely \emph{Eq.(\ref{helm})}, the Helmholtz
equation for conservation of vorticity.
\end{remark}

\begin{remark}
\label{EL}Of course, the main idea of this paper that we can get from the
Maxwell like equation that follows from the Einstein equation a Navier-Stokes
equation for an inviscid fluid depends on the existence of nontrivial
solutions of \emph{Eq.(\ref{f}), i.e., }$\mathring{F}_{0i}=(\boldsymbol{w}%
\times\boldsymbol{v})_{i}+d_{i}$. So, it is important to show that this
equation has nontrivial realizations for at least some Killing vector fields
living in $M$, such that $\langle M,\boldsymbol{g},D,\tau_{\boldsymbol{g}%
},\uparrow\rangle$ models a gravitational field. That this is the case, is
easily seem if we take, e.g., the Schwarzschild spacetime structure. As well
known \emph{\cite{mtw}} the vector field $\mathbf{A}=\partial_{\varphi
}=-\mathtt{x}^{2}\partial_{\mathtt{x}^{1}}+\mathtt{x}^{1}\partial
_{\mathtt{x}^{2}}$ is a Killing vector field for the Schwarzschild
metric\footnote{The spherical coordinate functions are $(r$,$\theta,\varphi
)$.}. The $1$-form field corresponding to it and living in Minkowski spacetime
is $\mathring{A}=\mathtt{x}^{2}d\mathtt{x}^{1}-\mathtt{x}^{1}d\mathtt{x}^{2}$.
Thus $\phi=0$ and $\boldsymbol{v}=(\mathtt{x}^{2},-\mathtt{x}^{1},0)$. This
gives $0=\mathring{F}_{0i}=-d_{i}+(\boldsymbol{w}\times\boldsymbol{v})_{i}%
,$i.e.,
\begin{equation}
\boldsymbol{d}=-\boldsymbol{v}\times(\nabla\times\boldsymbol{v)=}\nabla
\lbrack(\mathtt{x}^{1})^{2}+(\mathtt{x}^{2})^{2}], \label{nnn}%
\end{equation}
and \emph{Eq.(\ref{nse5}) }holds. More sophisticated examples for other
solutions of Einstein equations will be presented in another paper.
\end{remark}

\begin{remark}
For the example given in Remark \ref{EL} we have simply $A=f\mathring{A}$
where $f=r^{2}\cos^{2}\theta$. Thus in this case, we have the simple
expression%
\begin{equation}
F=df\wedge\mathring{A}+f\mathring{F}. \label{IMP}%
\end{equation}
There are many examples of Killing vector fields for which $A=f\mathring{A}$
and for such fields that developments given below in terms of $A$ are easily
translated in terms of $\mathring{A}$. In particular, when $A=f\mathring{A}$
we have the following identification of the components of the Lamb and
vorticity\ vector fields with the components of $F$,%
\begin{align}
l_{i}  &  =(\boldsymbol{w}\times\boldsymbol{v})_{i}=\frac{1}{f}F_{0i}-(d\ln
f\wedge\mathring{A})_{0i},\nonumber\\
\boldsymbol{w}_{i}  &  =-\frac{1}{2}%
{\textstyle\sum\nolimits_{i=1}^{3}}
\epsilon_{ijk}\left[  \frac{1}{f}F_{jk}-(d\ln f\wedge\mathring{A}%
)_{jk}\right]  . \label{lwF}%
\end{align}

\end{remark}

To continue, we recall that in order for the Navier-Stokes equation just
obtained to be compatible with Einstein equation it is necessary yet to take
into account Eq.(\ref{M1}), the non homogeneous Maxwell equation written in
the structure $\langle M,\boldsymbol{g},D,\tau_{\boldsymbol{g}},\uparrow
\rangle$, and the fact that $A$ is in the Lorenz gauge, namely
$\underset{\boldsymbol{g}}{\delta}A=0$, since these equations produce as we
are going to see constraints among the several fields involved, The
constraints involving the components of $A=\boldsymbol{g}(\mathbf{A,})$ with
$\mathbf{A}$ defined in Eq.(\ref{A}) are also encoded in Eq.(\ref{1}), which
\emph{must} now be expressed in terms of the objects defining the Minkowski
spacetime structure $\langle M=\mathbb{R}^{4},\boldsymbol{\mathring{g}%
},\mathring{D},\tau_{\boldsymbol{\mathring{g}}},\uparrow\rangle$.

Now, taking into account that $\mathring{D}\boldsymbol{\mathring{g}}=0$, we
have
\begin{equation}
D\boldsymbol{\mathring{g}}=\mathcal{A\in}\sec T_{0}^{2}M\otimes%
{\textstyle\bigwedge\nolimits^{1}}
T^{\ast}M, \label{amet}%
\end{equation}
where $\mathcal{A\in}\sec T_{0}^{2}M\otimes%
{\textstyle\bigwedge\nolimits^{1}}
T^{\ast}M$ is the non metricity tensor of $D$ with respect to
$\boldsymbol{\mathring{g}}$. In the coordinates $\langle\mathtt{x}^{\mu
}\rangle$ introduced above it follows that
\begin{equation}
\mathcal{A}=Q_{\alpha\beta\sigma}\boldsymbol{\gamma}^{\alpha}\otimes
\boldsymbol{\gamma}^{\beta}\otimes\boldsymbol{\gamma}^{\sigma}. \label{ameta}%
\end{equation}
Then, as it is well known\footnote{See, e.g., \cite{rodcap2007}.} the relation
between the coefficients $\Gamma_{\mu\alpha}^{\nu}$ and $\mathring{\Gamma
}_{\mu\alpha}^{\nu}$ associated to the connections $D$ and $\mathring{D}$
($D_{\boldsymbol{e}_{\mu}}\boldsymbol{\vartheta}^{\nu}=-\Gamma_{\mu\alpha
}^{\nu}\boldsymbol{\vartheta}^{\alpha}$,\ $\ \mathring{D}_{\boldsymbol{e}%
_{\mu}}\boldsymbol{\vartheta}^{\nu}=-\mathring{\Gamma}_{\mu\alpha}^{\nu
}\boldsymbol{\vartheta}^{\alpha}$) in an arbitrary coordinate vector
$\langle\frac{\partial}{\partial x^{\mu}}\rangle$ and covector $\langle
\boldsymbol{\vartheta}^{\nu}=dx^{\nu}\rangle$ bases --- associated to
arbitrary coordinate functions $\{x^{\mu}\}$ covering $U\subset M$ --- are
given by\footnote{We use that $\boldsymbol{\mathring{g}}=\mathring{g}_{\mu\nu
}\vartheta^{\mu}\otimes\vartheta^{\nu}=\mathring{g}^{\mu\nu}\vartheta_{\mu
}\otimes\vartheta_{\nu}$, where $\{\vartheta_{\mu}\}$ is the
\textit{reciprocal }basis of $\{\vartheta^{\mu}\}$, namely $\vartheta_{\mu
}=\mathring{g}_{\alpha\mu}\vartheta^{\alpha}$ and $\mathring{g}^{\mu\nu
}\mathring{g}_{\mu\kappa}=\delta_{\kappa}^{\nu}$. In the bases associated to
$\langle\mathtt{x}^{\mu}\rangle$ it is $\boldsymbol{\mathring{g}}=\eta_{\mu
\nu}\boldsymbol{\gamma}^{\mu}\otimes\boldsymbol{\gamma}^{\nu}=\eta^{\mu\nu
}\boldsymbol{\gamma}_{\mu}\otimes\boldsymbol{\gamma}_{\nu}$ \cite{jpa,aaca}.}
\begin{equation}
\Gamma_{\mu\alpha}^{\nu}=\mathring{\Gamma}_{\mu\alpha}^{\nu}+\frac{1}{2}%
S_{\mu\alpha}^{\nu}, \label{gamas}%
\end{equation}
where
\begin{equation}
S_{\alpha\beta}^{\rho}=\mathring{g}^{\rho\sigma}(Q_{\alpha\beta\sigma
}+Q_{\beta\sigma\alpha}-Q_{\sigma\alpha\beta}) \label{strain}%
\end{equation}
are the components of the so called \emph{strain tensor} of the connection.

In the coordinate bases $\langle\frac{\partial}{\partial\mathtt{x}^{\mu}%
}\rangle$ and $\langle\boldsymbol{\gamma}^{\mu}=d\mathtt{x}^{\mu}\rangle$,
associated to the coordinate functions $\langle\mathtt{x}^{\mu}\rangle$, it
follows that $\mathring{\Gamma}_{\mu\alpha}^{\nu}=0$ and in addition the
following relation for the Ricci tensor of $D$ holds:%
\begin{equation}
R_{\mu\nu}=J_{(\mu\nu)}.\nonumber
\end{equation}
Denoting $K_{\alpha\beta}^{\rho}=-\frac{1}{2}S_{\alpha\beta}^{\rho}$, the
$J_{(\mu\nu)}$ is the symmetric part of%
\begin{equation}
J_{\mu\alpha}=\mathring{D}_{\alpha}K_{\rho\mu}^{\rho}-\mathring{D}_{\rho
}K_{\alpha\mu}^{\rho}+K_{\alpha\sigma}^{\rho}K_{\rho\mu}^{\sigma}%
-K_{\rho\sigma}^{\rho}K_{\alpha\mu}^{\sigma}.\nonumber
\end{equation}
Now, if the Dirac operator associated to the Levi-Civita connection
$\mathring{D}$ of $\boldsymbol{\mathring{g}}$ is introduced by
\begin{equation}%
\bpartial
:=\boldsymbol{\vartheta}^{\mu}\mathring{D}_{\frac{\partial}{\partial x^{\mu}}%
}=\gamma^{\mu}\mathring{D}_{\frac{\partial}{\partial\mathtt{x}^{\mu}}}
\label{standard dirac}%
\end{equation}
it can be shown that\footnote{See Exercise 291 in \cite{rodcap2007}.}
\begin{equation}
\boldsymbol{\partial\wedge\partial}A=(%
\bpartial
\wedge%
\bpartial
\boldsymbol{)}\check{A}+\boldsymbol{L}^{\alpha}\underset{\boldsymbol{\mathring
{g}}}{\cdot}\gamma_{\alpha}\check{A}, \label{exerc291}%
\end{equation}
where $A=A_{\mu}\boldsymbol{\gamma}^{\mu}$, $\check{A}_{\kappa}:=\eta
_{\beta\kappa}g^{\beta\sigma}A_{\sigma}$, and $\boldsymbol{L}^{\alpha}%
=\eta^{\alpha\beta}J_{\beta\sigma}\boldsymbol{\gamma}^{\sigma}$. The symbol
$\underset{\boldsymbol{\mathring{g}}}{\cdot}$ denotes the scalar product
accomplished with $\boldsymbol{\mathring{g}}$. Since $%
\bpartial
\wedge%
\bpartial
\check{A}=\mathcal{\mathring{R}}^{\mathcal{\sigma}}\check{A}_{\sigma
}=\mathring{R}_{\alpha}^{\sigma}$ $\check{A}_{\sigma}\boldsymbol{\gamma
}^{\alpha}=0$ it reads
\begin{equation}
\boldsymbol{\partial\wedge\partial A}=\eta^{\alpha\beta}J_{\beta\alpha
}\boldsymbol{\check{A}}=\eta^{\alpha\beta}J_{\beta\alpha}\eta_{\iota\kappa
}g^{\iota\sigma}A_{\sigma}\boldsymbol{\gamma}^{\kappa}. \label{welldone}%
\end{equation}
According to Eq.(\ref{4}) $\boldsymbol{\partial\wedge\partial}%
A=\boldsymbol{\square}A$ \ and thus Eq.(\ref{1}) can also be written as
\[
\boldsymbol{\partial\wedge\partial}A=\frac{1}{2}RA+\boldsymbol{T}(A).
\]
Taking into account Eq.(\ref{welldone}), the following algebraic
equation\footnote{Of course, it is a partial differential equation that needs
to be satisfied by the components of the stress tensor of the connection.},
relating the components $A_{\sigma}$ to the components of the energy-momentum
tensor of matter and the components of the $\boldsymbol{g}$ field that is part
of the original LSTS, is obtained:
\begin{equation}
\eta^{\alpha\beta}J_{\beta\alpha}\eta_{\iota\kappa}g^{\iota\sigma}A_{\sigma
}=\frac{1}{2}g^{\mu\alpha}J_{(\mu\alpha)}A_{\kappa}+T_{\kappa}^{\sigma
}A_{\sigma}. \label{last}%
\end{equation}
Eq.(\ref{last}) are the constraints need to be satisfied by the variables of
our theory in order for the Navier-Stokes equation to be compatible with the
contents of Einstein equation.

As a last remark we observe that Eq.(\ref{last}) may be also interpreted as an
equation providing the energy-momentum tensor of the matter field as a
function of the variables entering the Navier-Stokes identification. \medskip

\section{Conclusions}

We demonstrated that for each Lorentzian spacetime representing a gravitation
field in General Relativity which contains an arbitrary\ Killing vector field
$\mathbf{A}$, the field $F=dA$ (where $A=\boldsymbol{g}(\mathbf{A},$ $)$)
satisfies Maxwell like equations with well determined current $1$-form field.
\ Moreover we showed that for all Killing $1$-form fields $\mathring
{A}=\boldsymbol{\mathring{g}}(\mathbf{A,)}$, when some identifications of the
components of $\mathring{A}$ and the variables entering the Navier-Stokes
equation are accomplished and in particular when the postulated nontrivial
condition (Eq.(\ref{f})--- $\mathring{F}_{0i}=(d\mathring{A})_{0i}=l_{i}%
-d_{i}$--- is satisfied, the Maxwell like equations for $\mathring{F}$ and
thus the ones for $F$ can be written as a Navier-Stokes equation representing
an \emph{inviscid fluid}. Thus, the Maxwell and Navier-Stokes like equations
found in this paper\footnote{In \cite{schm} a fluid satisfying a particular
Navier-Stokes equation is also shown to be approximately equivalent to
Einstein equation. Our approach is completely different from the one in
\cite{schm}.} are almost directly obtained from Einstein equation through
thoughtful identification of fields. All fields in our approach live in a
4-dimensional background spacetime, namely Minkowski spacetime and the
Lorentzian spacetime structures $\langle M,\boldsymbol{g},D,\tau
_{\boldsymbol{g}},\uparrow\rangle$ is considered only a (sometimes useful)
description of gravitational fields. Thus our approach is in contrast with the
very interesting and important studies in, e. g., \cite{breke,hub,rang} where
it is shown through some identifications that every solution for an
incompressible Navier-Stokes equation in a $(p+1)$-dimensional spacetime gives
rise to a solution of Einstein equation in $(p+2)$-dimensional spacetime. It
is worth also to quote \cite{pada} where it is also suggested an interesting
relation between Einstein equations and the Navier-Stokes equation. Finally we
remark that it is clear that we can find examples \cite{san} of Lorentzian
spacetimes that do not have any nontrivial Killing vector field. However, as
asserted in Weinberg \cite{wei}, all Lorentzian spacetimes that represent
gravitational fields of physical interest possess some Killing vector fields,
and we exhibit an example (Remark \ref{EL}) the approach of the present paper applies.

\section*{Acknowledgement}

{Authors are grateful to E. Capelas de Oliveira, J. F. T. Giglio, J. Vaz Jr.
and S. A. Wainer for discussions that motivate the present study. Finally, R.
da Rocha is grateful to CNPq grant 476580/2010-2, grant 304862/2009-6, and
grant 451042/2012-3 for financial support and F. G. Rodrigues is grateful to
CAPES for a Ph.D. fellowship. }

\appendix{}

\section{Some Mathematical Preliminaries}

In what follows $d$ and $\underset{\boldsymbol{g}}{\delta}$ denotes
respectively the differential and Hodge codifferential operators and if
$R\in\sec%
{\textstyle\bigwedge\nolimits^{r}}
T^{\ast}M$, $\delta R=(-1)^{r}\underset{\boldsymbol{g}}{\star}^{-1}%
d\underset{\boldsymbol{g}}{\star}R$ where $\underset{\boldsymbol{g}}{\star
}^{-1}$is the inverse of the Hodge star operator and
if\footnote{$\mathrm{sign}(\mathrm{\det}\boldsymbol{g})$ means the signal of
the deterninant of the matrix with entries $g_{\mu\nu}$.} $S\in\sec%
{\textstyle\bigwedge\nolimits^{s}}
T^{\ast}M$, $\underset{\boldsymbol{g}}{\star}^{-1}S=(-1)^{s(4-s)}%
\mathrm{sign}(\mathrm{\det}\boldsymbol{g)}\underset{\boldsymbol{g}}{\star}S.$
Of course, analogous formulas hold for $\underset{\boldsymbol{\mathring{g}%
}}{\delta}\mathcal{C}$.

If $\langle x^{\mu}\rangle$ are coordinates covering $U\subset M$ and
$\langle\boldsymbol{e}_{\mu}=\partial/\partial x^{\mu}\rangle$ a basis for
$TU$ and $\langle\boldsymbol{\vartheta}^{\mu}=dx^{\mu}\rangle$ the basis for
$T^{\ast}U$ dual to the basis $\langle\boldsymbol{e}_{\mu}\rangle$. Moreover
we call $\langle\boldsymbol{e}^{\mu}\rangle$ the reciprocal basis of
$\langle\boldsymbol{e}_{\mu}\rangle$ and $\langle\boldsymbol{\vartheta}_{\mu
}\rangle$ the reciprocal basis of $\langle\boldsymbol{\vartheta}^{\mu}\rangle
$. We have $\boldsymbol{g}(\boldsymbol{e}^{\mu},\boldsymbol{e}_{\nu}%
)=\delta_{\nu}^{\mu}$ and $\mathtt{g}(\boldsymbol{\vartheta}^{\mu
},\boldsymbol{\vartheta}_{\nu})=\delta_{\nu}^{\mu}$. Here $\mathtt{g}\in\sec
T_{0}^{2}M$ is the metric of the cotangent bundle such that if
$\boldsymbol{g=}g_{\mu\nu}\boldsymbol{\vartheta}^{\mu}\otimes
\boldsymbol{\vartheta}^{\nu}=g^{\mu\nu}\boldsymbol{\vartheta}_{\mu}%
\otimes\boldsymbol{\vartheta}_{\nu}$ then $\mathtt{g}=g^{\mu\nu}%
\boldsymbol{e}_{\mu}\otimes\boldsymbol{e}_{\nu}=g_{\mu\nu}\boldsymbol{e}^{\mu
}\otimes\boldsymbol{e}^{\nu}$.

We recall also that the Dirac operator\footnote{Of course given the structure
$\langle M,\boldsymbol{\mathring{g}},\mathring{D}\rangle$ we can define the
Dirac operator $\boldsymbol{\mathring{\partial}=\vartheta}^{\mu}\mathring
{D}_{\boldsymbol{e}_{\mu}}$which act on sections of the Clifford bundle
$\mathcal{C\ell(}M,\mathtt{\mathring{g}}).$} associated with the structure
$\langle M,\boldsymbol{g},D\rangle$ by:%
\begin{equation}
\boldsymbol{\partial=\vartheta}^{\mu}D\boldsymbol{_{\boldsymbol{e}_{\mu}}\ .}
\label{DIRAC}%
\end{equation}

A Dirac operator $\boldsymbol{\partial}$ act on sections of the Clifford
bundle of differential forms $\mathcal{C}\ell(M,\mathtt{g})$ and we have for
$C\in\sec\mathcal{C}\ell(M,\mathtt{g})$
\begin{equation}
\boldsymbol{\partial}C=\boldsymbol{\partial}\underset{\boldsymbol{g}%
}{\boldsymbol{\lrcorner}}C+\boldsymbol{\partial\wedge}C \label{dirac1}%
\end{equation}
where $\underset{\boldsymbol{g}}{\boldsymbol{\lrcorner}}$ denotes the left
contraction and $\wedge$ is the exterior product. It is a well known
\begin{equation}
\boldsymbol{\partial}\underset{\boldsymbol{g}}{\boldsymbol{\lrcorner}%
}C=-\underset{\boldsymbol{g}}{\delta}C\text{, \ \ }\boldsymbol{\partial\wedge
}C=dC, \label{dirac2}%
\end{equation}
and so%
\begin{equation}
\boldsymbol{\partial}=d\underset{\boldsymbol{g}}{-\delta} \label{dirac3}%
\end{equation}
Crucial for what follows is to recall that the square of the Dirac operator
$\boldsymbol{\partial}^{2}$ has tow non trivial decompositions, i.e.,
\begin{equation}
\boldsymbol{\partial}^{2}=-d\underset{\boldsymbol{g}}{\delta}%
-\underset{\boldsymbol{g}}{\delta}d=\boldsymbol{\partial}\wedge
\boldsymbol{\partial+\partial}\cdot\boldsymbol{\partial.} \label{dirac4}%
\end{equation}

Before continuing take notice $-d\underset{\boldsymbol{g}}{\delta}%
\neq\boldsymbol{\partial}\wedge\boldsymbol{\partial}$,
$-d\underset{\boldsymbol{g}}{\delta}\neq\boldsymbol{\partial}\cdot
\boldsymbol{\partial}$ and also $-\underset{\boldsymbol{g}}{\delta}%
d\neq\boldsymbol{\partial}\wedge\boldsymbol{\partial}$,
$-\underset{\boldsymbol{g}}{\delta}d\neq\boldsymbol{\partial}\cdot
\boldsymbol{\partial}$.

The operator $\square=\boldsymbol{\partial}\cdot\boldsymbol{\partial}$. is the
usual covariant D'Alembertian and $\boldsymbol{\partial}\wedge
\boldsymbol{\partial}$ is called the Ricci operator. The reason for that name
is that $(\boldsymbol{\partial}\wedge\boldsymbol{\partial)=\vartheta}^{\mu
}=\mathcal{R}^{\mu}:=R_{\nu}^{\mu}\boldsymbol{\vartheta}^{\nu}$ where $R_{\nu
}^{\mu}$ are the components of the Ricci tensor. The $1$-form fields where
$\mathcal{R}^{\mu}$ are called Ricci $1$-fields. We recall also that for any
$C\in\sec\mathcal{C}\ell(M,\mathtt{g})$ it is%
\begin{equation}
\underset{\boldsymbol{g}}{\star}C=\tilde{C}\boldsymbol{\tau}_{\boldsymbol{g}},
\label{dirac6}%
\end{equation}
where $\tilde{C}$ denotes the reverse of $C$. If $C=R\in\sec%
{\textstyle\bigwedge\nolimits^{r}}
T^{\ast}M\hookrightarrow\sec\mathcal{C}\ell(M,\mathtt{g})$ then $\tilde
{R}=(-1)^{\frac{1}{2}r(r-1)}R$. More details on the structures just introduced
can be found in \cite{rodcap2007}.

\section{A Comment on the Conserved Kommar Current}

Let $\mathbf{A}\in\sec TM$ be the generator of a one parameter group of
diffeomorphisms of $M$ in the spacetime structure $\langle M,\boldsymbol{g}%
,D,\tau_{\boldsymbol{g}},\uparrow\rangle$ which is a model of a gravitational
field generated by $\mathbf{T}\in\sec T_{0}^{2}M$ (the matter plus\ non
gravitational fields energy-momentum momentum tensor) in Einstein GR. It is
quite obvious that if we define $F=dA$, where $A=\boldsymbol{g}(\mathbf{A,}%
)\in\sec%
{\textstyle\bigwedge\nolimits^{1}}
T^{\ast}M$ \ then the current
\begin{equation}
J_{\boldsymbol{K}}=-\underset{\boldsymbol{g}}{\delta}F\label{n0}%
\end{equation}
is conserved, i.e.,
\begin{equation}
\underset{\boldsymbol{g}}{\delta}J_{\boldsymbol{K}}=0.\label{n01}%
\end{equation}
Surprisingly such a trivial \emph{mathematical} result seems to be very
important by people working in GR who call $J_{\boldsymbol{K}}$. the Kommar
current\footnote{Komar called a \emph{related} quantity the generalized flux.}
\cite{kommar,komar}. Komar called\footnote{$V$ denotes a spacelike
hypersurface and $S=\partial V$ \ its boundary. Usualy the integral
$\mathfrak{E}$ is calculated at a constant $x^{0}$ time hypersurface and the
limit is taken for $S$ being the boundary at infinity.}
\begin{equation}
\mathfrak{E}=-%
{\textstyle\int\nolimits_{V}}
\underset{\boldsymbol{g}}{\star}J_{\boldsymbol{K}}=%
{\textstyle\int\nolimits_{\partial V}}
\underset{\boldsymbol{g}}{\star}F\label{n02}%
\end{equation}
the \emph{generalized energy.}

To understand why $J_{\boldsymbol{K}}$ is considered important write the
action for the gravitational plus matter and non gravitational fields as
\begin{equation}
\mathcal{A=}%
{\textstyle\int}
\mathcal{L}_{g}+%
{\textstyle\int}
\mathcal{L}_{m}=-\frac{1}{2}%
{\textstyle\int}
R\boldsymbol{\tau}_{\boldsymbol{g}}+%
{\textstyle\int}
\mathcal{L}_{m}.\label{n022}%
\end{equation}

Now, under the (infinitesimal) diffeormorphism $h:M\rightarrow M$ generated by
$\mathbf{A}$ we have that $\boldsymbol{g}\mapsto\boldsymbol{g}^{\prime
}=h^{\ast}\boldsymbol{g=g+\delta g}$ where\footnote{Please do not confuse
$\boldsymbol{\delta}$ with $\underset{\boldsymbol{g}}{\delta}$} the variation
$\boldsymbol{\delta g}=-\pounds _{\mathbf{A}}\boldsymbol{g}$ and taking into
account Cartan's magical formula ($\pounds _{\mathbf{A}}J_{\boldsymbol{K}%
}=A\underset{\boldsymbol{g}}{\boldsymbol{\lrcorner}}dJ_{\boldsymbol{K}%
}+d(A\underset{\boldsymbol{g}}{\boldsymbol{\lrcorner}}J_{\boldsymbol{K}})$,
$J_{\boldsymbol{K}}\in\sec%
{\textstyle\bigwedge}
T^{\ast}M$) we have
\begin{align}
\boldsymbol{\delta}\mathcal{A}  &  =%
{\textstyle\int}
\boldsymbol{\delta}\mathcal{L}_{g}+%
{\textstyle\int}
\boldsymbol{\delta}\mathcal{L}_{m}\nonumber\\
&  =-%
{\textstyle\int}
\pounds _{\mathbf{A}}\mathcal{L}_{g}-%
{\textstyle\int_{\mathbf{A}}}
\pounds \mathcal{L}_{m}\nonumber\\
&  =-%
{\textstyle\int}
d(A\lrcorner\mathcal{L}_{g})-%
{\textstyle\int}
d(A\lrcorner\mathcal{L}_{m})\nonumber\\
&  :=%
{\textstyle\int}
d(\underset{\boldsymbol{g}}{\star}\mathcal{C)} \label{n03}%
\end{align}

Of course, if $\underset{\boldsymbol{g}}{\star}\mathcal{C}$ is null on the
boundary of the integration region we have $\boldsymbol{\delta}\mathcal{A}=0$.
On the other hand if we write \cite{landau}
\begin{equation}
\mathcal{A=-}\frac{1}{2}%
{\textstyle\int}
R\sqrt{-\det\boldsymbol{g}}dx^{0}dx^{1}dx^{2}dx^{3}+%
{\textstyle\int}
L_{m}\sqrt{-\det\boldsymbol{g}}dx^{0}dx^{1}dx^{2}dx^{3}\label{n04}%
\end{equation}
we have (writing $\mathcal{E}^{\mu}:=\mathcal{G}^{\mu}-\mathcal{T}^{\mu}$
where $\mathcal{G}^{\mu}:=\mathcal{R}^{\mu}-\frac{1}{2}R\boldsymbol{\vartheta
}^{\mu}$ are the Einstein $1$-form fields\footnote{$\mathcal{G}^{\mu}=G_{\nu
}^{\mu}\boldsymbol{\vartheta}^{\nu}$ where $G_{\nu}^{\mu}=R_{\nu}^{\mu}%
-\frac{1}{2}\delta_{\nu}^{\mu}$ are the components of the Einstein tensor.}
and $\mathcal{T}^{\mu}=T_{\nu}^{\mu}\boldsymbol{\vartheta}^{\nu}$ the
energy-momentum $1$-form fields\footnote{The energy momentum tensor is
$\mathbf{T=}T_{\upsilon}^{\mu}\boldsymbol{\vartheta}^{\nu}\otimes
\boldsymbol{e}_{\mu}$.} and recalling that $D_{\mu}G^{\mu\nu}=0=D_{\mu}%
T^{\mu\nu}$)
\begin{align}
\boldsymbol{\delta}\mathcal{A} &  =\mathcal{-}\frac{1}{2}%
{\textstyle\int}
E^{\mu\nu}(\mathfrak{L}_{\mathbf{A}}\boldsymbol{g})_{\mu\nu}\sqrt
{-\det\boldsymbol{g}}dx^{0}dx^{1}dx^{2}dx^{3}\nonumber\\
&  =-%
{\textstyle\int}
E^{\mu\nu}D_{\mu}A_{\nu}\sqrt{-\det\boldsymbol{g}}dx^{0}dx^{1}dx^{2}%
dx^{3}\nonumber\\
&  =-%
{\textstyle\int}
D_{\mu}(E^{\mu\nu}A_{\nu})\sqrt{-\det\boldsymbol{g}}dx^{0}dx^{1}dx^{2}%
dx^{3}\nonumber\\
&  =-%
{\textstyle\int}
(\boldsymbol{\partial}\underset{\boldsymbol{g}}{\boldsymbol{\lrcorner}%
}\mathcal{E}^{\nu}A_{\nu})\boldsymbol{\tau}_{\boldsymbol{g}}\nonumber\\
&  =%
{\textstyle\int}
\underset{\boldsymbol{g}}{\star}\underset{\boldsymbol{g}}{\delta}%
(\mathcal{E}^{\nu}A_{\nu})\nonumber\\
&  =-%
{\textstyle\int}
\underset{\boldsymbol{g}}{d\star}(\mathcal{E}^{\nu}A_{\nu})\label{n05}%
\end{align}

From Eqs.(\ref{n03}) and (\ref{n04}) we have immediately that
\begin{equation}%
{\textstyle\int}
d(\underset{\boldsymbol{g}}{\star}\mathcal{E}^{\nu}A_{\nu}%
)+d(\underset{\boldsymbol{g}}{\star}\mathcal{C)}=0,\label{n06}%
\end{equation}
and thus
\begin{equation}
\underset{\boldsymbol{g}}{\delta}(\mathcal{E}^{\nu}A_{\nu}%
)+\underset{\boldsymbol{g}}{\delta}\mathcal{C}=0\label{n07}%
\end{equation}
Thus the current $\mathcal{C}\in\sec%
{\textstyle\bigwedge\nolimits^{1}}
T^{\ast}M$ is conserved if the field equations $\mathcal{E}^{\nu}=0$ are
satisfied. An equation (in component form) equivalent to Eq.(\ref{n07})
already appears in \cite{kommar} (and also previously in \cite{berg} ) who
took $\mathcal{C=E}^{\nu}A_{\nu}+N$ where $\underset{\boldsymbol{g}}{\delta
}N=0$.

Here to continue we prefer to write an identity involving only
$\boldsymbol{\delta}\mathcal{A}_{g}=%
{\textstyle\int}
\boldsymbol{\delta}\mathcal{L}_{g}$. Proceeding exactly as before we get
putting $\mathcal{G(}A\mathcal{)=G}^{\mu}A_{\mu}$ there exists $\mathcal{P}%
\in\sec%
{\textstyle\bigwedge\nolimits^{1}}
T^{\ast}M$ such that%
\begin{equation}
\boldsymbol{\partial\lrcorner}\mathcal{G(}A\mathcal{)+}\boldsymbol{\partial
\lrcorner}\mathcal{P}=0. \label{n1}%
\end{equation}
and we see that we can identify%
\begin{equation}
\mathcal{P}:=-\mathcal{G}^{\mu}A_{\mu}+L \label{n4}%
\end{equation}
where $\underset{\boldsymbol{g}}{\delta}L=0$. Now, we claim that we can find
$L\in\sec%
{\textstyle\bigwedge\nolimits^{1}}
T^{\ast}M$ such that $\mathcal{P}=-\mathcal{G}^{\mu}A_{\mu}+L=\delta dA$. Let
us find $L$ and investigate if we can give some nontrivial physical meaning to
such $\mathcal{P}\in\sec%
{\textstyle\bigwedge\nolimits^{1}}
T^{\ast}M$

In order to prove our claim we observe that we can write
\begin{align}
\mathcal{G}^{\mu}A_{\mu}  &  =\mathcal{R}^{\mu}A_{\mu}-\frac{1}{2}%
RA\nonumber\\
&  =\boldsymbol{\partial}\wedge\boldsymbol{\partial}A-\frac{1}{2}RA\nonumber\\
&  =\boldsymbol{\partial}\wedge\boldsymbol{\partial}A+\boldsymbol{\partial
}\cdot\boldsymbol{\partial}A-\frac{1}{2}RA-\boldsymbol{\partial}%
\cdot\boldsymbol{\partial}A\nonumber\\
&  =\boldsymbol{\partial}^{2}A-\frac{1}{2}RA-\boldsymbol{\partial}%
\cdot\boldsymbol{\partial}A\nonumber\\
&  =-\underset{\boldsymbol{g}}{\delta}dA-d\underset{\boldsymbol{g}}{\delta
}A-\frac{1}{2}RA-\boldsymbol{\partial}\cdot\boldsymbol{\partial}A \label{n5}%
\end{align}
Then we take%
\begin{equation}
\mathcal{P}:=-\mathcal{G}^{\mu}A_{\mu}-d\underset{\boldsymbol{g}}{\delta
}A-\frac{1}{2}RA-\boldsymbol{\partial}\cdot\boldsymbol{\partial}A=\text{
}\underset{\boldsymbol{g}}{\delta}dA \label{N6}%
\end{equation}
and of course\footnote{Note that since $\underset{\boldsymbol{g}}{\delta
}(\mathcal{G}^{\mu}K_{\mu})=0$ it follows from Eq.(\ref{n6aa}) that indeed
$\underset{\boldsymbol{g}}{\delta}L=0$.}%
\begin{equation}
L=-d\underset{\boldsymbol{g}}{\delta}A-\frac{1}{2}RA-\boldsymbol{\partial
}\cdot\boldsymbol{\partial}A \label{n6aa}%
\end{equation}

Thus since $\mathcal{G}^{\mu}A_{\mu}=\mathcal{T}^{\mu}A_{\mu}:=\boldsymbol{T}%
\mathcal{(}A\mathcal{)}$ we have%

\begin{equation}
\delta dA=-\boldsymbol{T}\mathcal{(}A\mathcal{)}-d\underset{\boldsymbol{g}%
}{\delta}A-\frac{1}{2}RA-\boldsymbol{\partial}\cdot\boldsymbol{\partial}A
\label{n6a}%
\end{equation}
We can write Eq.(\ref{n6a}) taking into account that $R=-T=-T_{\mu}^{\mu}$ and
putting $F:=dA$ as
\begin{equation}
\underset{\boldsymbol{g}}{\delta}F=-J_{\boldsymbol{K}} \label{the equation}%
\end{equation}%
\begin{equation}
J_{\boldsymbol{K}}=\boldsymbol{T}(A)-\frac{1}{2}TA\mathcal{+}\text{ }d\delta
A+\boldsymbol{\partial}\cdot\boldsymbol{\partial}A \label{kocurr}%
\end{equation}
Eq.(\ref{kocurr}) gives the explicit form for the Kommar
current\footnote{Something that is not given in \cite{kommar}}. Moreover,
since $\underset{\boldsymbol{g}}{\delta}F=\star d\star F$
\begin{align}
d\star F  &  =\mathcal{\star}^{-1}\left(  \boldsymbol{T}(A)-\text{ }\frac
{1}{2}TA\mathcal{+}\text{ }d\underset{\boldsymbol{g}}{\delta}%
A+\boldsymbol{\partial}\cdot\boldsymbol{\partial}A\right) \label{n9}\\
&  =\mathcal{\star}\left(  -\boldsymbol{T}(A)+\text{ }\frac{1}{2}%
TK\mathcal{-}\text{ }d\underset{\boldsymbol{g}}{\delta}A-\boldsymbol{\partial
}\cdot\boldsymbol{\partial}A\right)
\end{align}
and thus taking into account Stokes theorem
\[%
{\textstyle\int\nolimits_{V}}
d\star F=%
{\textstyle\int\nolimits_{\partial V}}
\star F
\]
we arrive at the conclusion that the quantity
\begin{align}
\mathcal{E}  &  :=-\frac{1}{8\pi}\int_{S=\partial V}\star F\label{n10}\\
&  =\frac{1}{8\pi}%
{\textstyle\int\nolimits_{V}}
\mathcal{\star}\left(  \mathcal{(}\boldsymbol{T}(A)-\frac{1}{2}TA\mathcal{+}%
\text{ }d\underset{\boldsymbol{g}}{\delta}A+\boldsymbol{\partial}%
\cdot\boldsymbol{\partial}A\right)  \label{n11}%
\end{align}

\begin{remark}
As we already remarked an equation equivalent to \emph{Eq.(\ref{n10})} has
already been obtained in\emph{ \cite{kommar,komar}} who called that quantity
the conserved \emph{generalized energy.} But according to our best knowledge
\emph{Eq.(\ref{n11})} is new and it shows explicitly that all terms in the
integrand are legitimate $3$-form fields and thus the value of the integral
is, of course, independent of the coordinate system used to calculate it.

However, considering that for each $\mathbf{A}$ $\in\sec TM$ that generates a
one parameter group of diffeomorphisms of $M$ we have a conserved quantity it
is not in our opinion appropriate to think in this quantity as a generalized
energy. Indeed, why should the energy depends on terms like
$d\underset{\boldsymbol{g}}{\delta}A$ and $\boldsymbol{\partial}%
\cdot\boldsymbol{\partial}A$ if $A$ is not a dynamical field?

We know from \emph{Eq.(\ref{1}) }that when $\mathbf{A}$ is a Killing vector
field the term $d\underset{\boldsymbol{g}}{\delta}A=0$ and
$\boldsymbol{\partial}\cdot\boldsymbol{\partial}A=\boldsymbol{T}%
\mathcal{(}A\mathcal{)-}\frac{T}{2}A$ and thus \emph{Eq.(\ref{n11}) reads}%
\begin{equation}
\mathcal{E}=\frac{1}{8\pi}%
{\textstyle\int\nolimits_{V}}
\mathcal{\star(}\boldsymbol{T}(A)-\frac{1}{2}TA) \label{n12}%
\end{equation}
which is a conserved quantity\emph{\footnote{Observe that when $\mathbf{A}$ is
a Killing vector field the quantities $%
{\textstyle\int\nolimits_{V}}
\star\boldsymbol{T}(A)$ and \ $%
{\textstyle\int\nolimits_{V}}
\frac{1}{2}\star TA$ are separately conserved as it is easily verified.}}
which at least is directly associated with the energy-momentum tensor of the
matter plus non gravitational fields\emph{\footnote{An equivalent formula
appears, e.g., \ as Eq.(11.2.10) in \cite{wald}. However, it is to be
emphasized here the simplicity and trasnparency of our approach concerning
traditional ones based on classical tensor calculus.}}. For a Schwarzschild
spacetime, as well known, $\mathbf{A=\partial/\partial}t$ is a timelike
Killing vector field and in his case since the components of $\mathbf{T}$ are
$T_{\nu}^{\mu}=\frac{4\pi}{\sqrt{-\det\boldsymbol{g}}}\rho(r)v^{\mu}v_{\nu}$
and $v^{i}v_{j}=0$ \emph{(}since $v^{\mu}=\frac{1}{\sqrt{g_{00}}}%
(1,0,0,0)$\emph{)} we get $\mathcal{E}=m$.
\end{remark}

\begin{remark}
Originally Kommar obtained the same result directly from \emph{Eq.(\ref{n11})
supposing that the generator of the one parameter group of diffeomorphism was
}$\mathbf{A}=\partial/\partial t$, so he got $\mathcal{E}=m$ by pure chance.
Had he \ picked another vector field generator of a one parameter group of
diffeomorphisms $\mathbf{A\neq\partial/\partial}t$, he of course, would not
obtained that result.
\end{remark}

\begin{remark}
The previous remark shows clearly that the above approach does not to solve
the energy-momentum conservation problem for the system consisting of the
matter plus non gravitational fields plus the gravitational field. It only
gives a conserved energy for the matter plus non gravitational fields if the
spacetime structure possess a timelike Killing vector field. To claim that a
solution for total energy-momentum of the total system\emph{\footnote{The
total system is the system consisting of the gravitational plus matter and non
gravitational fields.}} problem exist it is necessary to find a way to define
a total energy-momentum $1$-form for the total system. This can only be done
if the spacetime structure modelling a gravitational field (generated by the
matter plus non gravitational fields energy-momentum tensor $\mathbf{T}%
$\emph{)} possess additional structure, or if we interpret \ the gravitational
field as a field in the Faraday sense living in Minkowski spacetime as
discussed in\emph{ \cite{fr2010,rod2011}}.
\end{remark}

\begin{remark}
More important is to take in mind that \emph{Eq.(\ref{the equation}) }where
$F=dA\ $where $A=\boldsymbol{g}(\mathbf{A,})$ with $\mathbf{A}$ $\in\sec TM$
an arbitrary generator of a one parameter group of diffeomorphisms of $M$
(part of the structure $\langle M,\boldsymbol{g},D,\mathbf{\tau}%
_{\boldsymbol{g}}\boldsymbol{,\uparrow\rangle}$ encodes Einstein equation.
However, since it is only when $\mathbf{A}$ is a Killing vector field that the
Maxwell like equation \emph{Eq.(\ref{the equation})} seems to have at least
some physical meaning \ in Section 2 we will derive some of the above results
again \emph{(}with a different and simple method\emph{)} when $\mathbf{A}$ is
a Killing vector field. It will be the Maxwell like equation for this case
that will be shown to give place to a Navier-Stokes \emph{(}like\emph{)}
equation for an inviscid fluid.
\end{remark}

\end{document}